\documentstyle [12pt]{article}
\begin {document}
\textheight=20cm
\textwidth=15cm
\voffset=-1.2cm
\hoffset=-0.5cm
\baselineskip25pt
\parskip16pt
\def\beq{\begin {array}}
\def\eeq{\end {array}}

\title{The Quark Gluon Pion Plasma}
\author{
Vikram Soni \footnote {\noindent vsoni@del3.vsnl.net.in}
\\ {\it National Physical Laboratory, New Delhi, India}
 \\
\and  Moninder S. Modgil  \footnote {\noindent
 msingh@iitk.ac.in}  Deshdeep Sahdev \footnote{\noindent
 ds@iitk.ac.in}
 \\{\it Department of Physics,} \\{\it Indian Institute 
 of Technology, Kanpur, 208016},\\ {\it India}}

\maketitle

\begin{abstract} 
While it is commonly believed that there is a {\it direct}
transition from the hadronic to a quark gluon phase at high
temperature, it would be prejudicial to rule out a sequence 
of dynamically generated intermediate scales. 
Using as guide, an effective lagrangian
with unconfined gluons and constituent quarks, interacting 
with a chiral multiplet, 
we examine a scenario in which the system undergoes
first-order transitions at $ T_{comp}$, the compositeness 
scale of the pions, at $T_{\chi}$, the scale for 
spontaneous chiral symmetry breaking, and 
at  $T_c$, the confinement temperature. 

We find that at current energies, it is likely that the 
formation temperature
of the plasma, $ T_0 < T_{comp} $, and that this is therefore 
a quark gluon pion plasma (QGPP) rather than
the usual quark gluon plasma (QGP).
We propose some dilepton-related signatures of this scenario. 
\end{abstract}

We know that quarks and gluons are confined as hadrons and chiral 
symmetry is spontaneously broken. We do not however know if 
these two phenomena set in at an identical temperature.
It is unlikely that the chiral symmetry restoration 
energy/temperature scale is {\it lower} than that of confinement,
because if it were, hadrons would show parity doubling below the 
confinement but above the chiral breaking scale. This is not seen
either experimentally or in finite temperature lattice
simulations. Determining whether it is {\it higher} even
in the simplest case of 
QCD with a two flavour $SU_2(L)\times SU_2(R)$ chiral symmetry 
is not straight-forward. Indeed, while there is a bonafide order parameter 
for chiral symmetry breaking --- namely, the mass of the constituent 
quark --- the Wilson loop in the presence of dynamical quarks,
is {\it not} a valid order parameter for confinement ---
and there are no other known candidates.
However, by looking at energy density or specific heat, we can get a
fair idea of the change in the number of operational degrees of freedom
or particle modes with temperature. The relevant lattice calculations 
indicate that the drop from the
large number of degrees of freedom in the QGP/QGPP phase to a
few degrees
of freedom in the hadronic one takes place in one broad step in
temperature, suggesting that the two transitions are close
but not necessarily identical, i.e. that 
$T_{\chi} \stackrel{\stackrel{}{>}}{\stackrel{{\sim}}{}} T_c$.

In the same spirit, we could ask whether pions necessarily come
apart the moment chiral symmetry is restored, or equivalently,
whether the compositeness scale for the pion, $T_{comp}$, 
coincides with $T_{\chi}$. The following analogy with 
superconductivity (SC) leads us to conclude that it may not:

In the usual BCS theory, the formation of Cooper pairs at $T_{crit}$
coincides with the appearance of a SC order parameter. This corresponds
to $T_{comp} = T_{crit}$, as there are no Cooper pairs at scales 
above $T_{crit}$. Here, $T_{comp}$ is the compositeness scale for 
Cooper pairs, which are normally described as undergoing 
momentum space pairing, the interelectron distance being much
less than the size of a Cooper pair. This is indeed what happens
for a weak pairing interaction.
For a much stronger interaction, Cooper pairs may actually
form by real space pairing. The Cooper pair size is then smaller than the 
interelectron distance and it is possible that, even when SC is lost, 
i.e. $ T > T_{crit}$, pairs continue to exist simply as bound states. 

Evidence for the related precursor phenomena for the chiral transition 
had been suggested long ago $ \cite {ref12}$. Recently, there have been 
several calculations of the QCD counterpart of the non-condensed, paired,
pseudogap phase of high $ T_{c} $ superconductivity $ \cite{ref13}$,
which suggest that the strong pairing situation is likely for the strong
interactions.

We shall accordingly take $T_{comp} > T_{\chi}$, and argue that
heavy-ion collisions are the ideal platform for testing this 
assumption.
In doing so, we shall use the following $SU_2(L) \times
SU_2(R)$
effective lagrangian, in which the quarks are coupled to a 
chiral multiplet, $[\pi, \sigma]$ \cite {ref2,ref3,ref4},
to keep track of particle modes at various scales: 

\begin{eqnarray}
 L = - \frac{1}{4} G^a_{\mu v} G^a_{\mu v}
 - \sum {\overline{\psi}} \left( D + g_y(\sigma +
i\gamma_5 \vec \tau \vec \pi)\right) \psi \nonumber \\
- \frac{1}{2} (\partial_
\mu \sigma)^2 - \frac{1}{2} (\partial_\mu \vec \pi)^2 
 - \frac{1}{2} \mu^2 (\sigma^2 + \vec \pi^2) - \frac{\lambda}{4}
(\sigma^2 + \vec \pi^2)^2 + \hbox{const}
\end{eqnarray}

This has chiral symmetry-breaking but no confinement. The 
masses of the scalar/pseudoscalars and fermions, obtained by
minimizing the potentials above, are $\mu^2=-\lambda<\sigma>^2$
and $m^2_\sigma=2\lambda<\sigma>^2$ respectively.
This theory reproduces several broad features of the strong
interaction at the mean field level \cite {ref5,ref6}, 
indicating that it is, perhpas, a reasonable guide to 
the physics at intermediate scales. More specifically,  
\begin{enumerate}
\item It provides a nucleon, which is realized as a soliton 
with quarks 
bound in a skyrmion configuration \cite {ref3,ref5}. 
Such a nucleon:
\begin{enumerate}
\item Gives a natural explanation for the `Proton spin 
puzzle':
Quarks in the background fields are in a spin, `isospin'
singlet state in which the quark spin operator averages to zero. If we
collectively quantize the soliton to get states of good spin and
isospin, the quark spin operator picks up a small, non-zero 
expectation value \cite {ref7}.
\item Seems to naturally produce the Gottfried sum rule \cite {ref8}.
\item Yields, from first principles ({\it albeit} with some
drastic QCD evolution), a set of structure functions for the 
nucleon which are close to the experimental ones \cite {ref9}.
\end{enumerate}
\item As a finite temperature field theory, it yields screening masses 
that match with those obtained from the lattice simulation of
finite temperature QCD
with dynamical quarks \cite {ref10}.
\item It gives a consistent equation of state for strongly
interacting matter at all densities  \cite {ref11,ref5}.
\end{enumerate}
This lagrangian and the above discussion on 
temperature scales lead us to consider the following scenario:
\begin{enumerate}
\item For $ T > T_{comp}$, we have  gluons and 
massless current quarks, in a pure QGP phase.
\item For $ T_{comp} > T > T_{\chi}$, we have a QGPP phase for
which the particle mode count
is (i) 16 modes from 8 massless gluons, (ii) 24 from 2 flavors 
of massless current quarks and anti-quarks, and (iii) 3 
degenerate pions, which are present, in this
temperature range, in the form of non-Goldstone-boson bound 
states, which we {\it assume} to be massless. (We 
neglect the $\sigma$, which would otherwise contribute
one more mode).

The operational degrees of freedom in this phase are then 
40 --- three more than in the QGP phase. This increase
may be visible as a slight bump on an `Energy Density' {\it vs}
`Temperature' plot, in a high resolution lattice simulation.
\item For $T_{\chi} > T > T_c$, 
chiral symmetry is spontaneously broken 
and the degrees of freedom are
(i) gluons, (ii) constituent quarks, which
acquire a constituent mass of $300-400$ MeV from the
spontaneous breaking of $\chi$-symmetry and are,
as a result, Boltzman suppressed given that $T_{\chi} \simeq
T_c$ is expected \cite {ref1} to be $O(200 MeV)$, 
(iii) pions as Goldstone bosons, and 
(iv) the $\sigma $, which also acquires a mass and may
be neglected.

The effective degrees of freedom are thus 19, at this stage.
\item For $T_c > T$, quarks are confined, and the only
massless modes left are the 3 Goldstone bosons/pions.
\end{enumerate}
We can get an estimate of $T_{comp}$ by examining the 
behaviour of the above Lagrangian as a function of energy. 
We find that at a certain scale, both the
wave function renormalization constant, $Z$, for the $\pi$ and $\sigma$,
{\it and} the quartic scalar interaction, vanish, leaving us with
a Yukawa and a scalar mass term. The vanishing of $Z$, or 
equivalently, of the kinetic
term for the scalar and pseudoscalars, means that these are no longer dynamical
degrees of freedom, i.e. they have ceased to exist as 
composite entities. Eliminating them, via their field 
equation,
leaves behind  a  four-fermi term which
gets weaker with increasing energy. While these results, 
gleaned
using perturbative RNG, cannot be used all the way out to  
where Z goes to zero (since the Yukawa coupling has a Landau 
singlarity there), they can perhaps give a reasonable
estimate for the compositeness scale. This works out to be
around $700-800$ MeV \cite {vs}, well above 
the temperatures attained in the present-day heavy ion 
colliders. 
This implies that the initial state produced by these
is very probably a QGPP.

To investigate this possibility further, we adapt to our 
scenario, the considerations of \cite {ref1}, 
which gives a very transparent and simple treatment of the 
underlying physics. In particular, following \cite {ref1}, 
we treat each of our transitions
as first order, with each producing a mixed phase in which 
the number of degrees of freedom changes continuously, as one 
phase gives way to the other. 

A plasma which thermalizes at a temperature, $T_0$, where
$T_{comp} < T_0 < T_{\chi}$, a proper time, $t_0$, after 
nuclear impact, then evolves as follows:
\begin{enumerate}
\item It Bjorken expands and cools \cite {bjorken}, reaching
$T_{\chi}$ at time, $t_{x1} = (T_0/T_{\chi})^3 t_0$.
\item At $ T_{\chi}$, it undergoes a first-order transition
from the quark-gluon-pion to the chirally broken, gluon-pion
phase, from
$t_{x1}$ to $r_1 t_{x1}$, where $r_1 = 40/19$ is the ratio 
of the degrees of freedom in the initial  
and final phases.

N.B. If we assume that chiral restoration and deconfinement 
occur simultaneously at $T_{\chi} = T_c$, the quark phase 
lasts in the mixture from time $t_{1}=t_{x1}$ to $t_{2} = r t_{1}$ 
(where $r = 37/3$), which is substantially longer.
\item Having passed to the chirally-broken phase,
the plasma undergoes a second Bjorken expansion and thereby 
cools from $T_{\chi}$ to $T_c$. 
Since $T_{\chi} \simeq T_c$, this part of the 
evolution may, however, be neglected.
\item At $T_c$, we get a mixture of  
the gluon-pion and the purely hadronic confined
phases, which lasts from $ r_1 t_{x1} $ to  $t_{x2} =
r_2 r_1 t_{x1}$, where $r_2$ is $19/3$.
\item Finally the pion phase expands from $T_c$ to $T_f$, the
freezeout temperature \cite {ref1}, at which pions loose
thermal contact with one another.
This phase begins at time $ t_{x2}$ and ends at 
$t_f=( T_c /T_f)^3 t_{x2} $.
\end{enumerate}
The basic changes in this scenario when compared to that 
of \cite {ref1} are
that, (i) the pionic phase starts much earlier, at the 
initial temperature, 
$T_0$, and not after confinement at $T_c$ and (ii) the
quark phase is suppressed above $T_c$ and is thus shortened.

To see how these changes are reflected in the number of 
dileptons emitted by the plasma, 
we note that
these come from $\pi^+\pi^-$ and $q\bar{q}$ annihilations.
The  cross-section for $q \overline{q} \rightarrow 
l^{+}l^{-}$, summed over the spin,
flavour and color of all quarks is 
\begin {eqnarray}
\sigma_q[M]=F_q\frac{4
\pi}{3}(\frac{\alpha}{M})^2\sqrt{1-4(\frac{m_l}{M})^2}
\bigg(1+2(\frac{m_l}{M})^2\bigg)
\end {eqnarray}
where the numerical factor, 
$F_q=20/3$, if we include just the $u$ and $d$ quarks in the
flavor sum.

The cross-section for $\pi^{+}\pi^{-}\rightarrow l^{+}l^{-}$ 
likewise is,
\begin {eqnarray}
\sigma_\pi[M]=F_\pi[M] (4 \pi/3) (\alpha/M)^2 \sqrt{1-4(m_l/M)^2}
 (1+2 (m_l/M)^2)\sqrt{1-4 (m_\pi/M)^2}
\end {eqnarray}
where the pion form factor is given by,
\begin {eqnarray}
F_\pi[M]=m_\rho^4/((m_\rho^2-M^2)^2+m_\rho^2 \Gamma_\rho^2)
+(1/4) m_{\rho'}^4/((m_{\rho'}^2-M^2)^2+m_{\rho'}^2 \Gamma_{\rho'}^2)
\end {eqnarray}
The masses and decay widths of $\rho$ and $\rho'$ resonances 
are $m_\rho=775$ MeV, $m_{\rho'}=1.6$ GeV, $\Gamma_\rho=155$ MeV, 
$\Gamma_{\rho'}=260$ MeV respectively.

The rate for producing pairs, per unit space-time, with 
invariant mass squared, 
$M^2$, and transverse energy, $E_T=\sqrt{p_T^2+M^2}$, where
$p_T$ is the momentum transverse to the beam axis, is
(for $a = q, \pi$),
\begin{equation}
\frac{d N_a}{d^4xdM^2dE_T} =   
\frac{\sigma_a[M]M^2}{4(2\pi)^4}(1-4\frac{m_a^2}{M^2})
E_T K_0(E_T/T)=
A_a[M]E_T K_0(E_T/T)
\end{equation}
While using Bjorken's model, it is expedient to write 
$d^4x = d^2x_Tdytdt$ where
$t$ is the proper time and $y$, the rapidity, of the fluid
element. For central collisions of equal mass nucleii
$d^2x_T = \pi R_A^2$, where $R_A$ is the nuclear radius.
We can further perform the $t$-integral to get:
\begin{eqnarray}
\label{qXsctn}
\frac{d N^{MT}_q}{dydM^2dE_T} = \pi 
R_A^2 A_q(M)  
\bigg\{3 E_T^{-5} T_0^6 t_0^2 (G[E_T/T_0]-G[E_T/T_c])
\nonumber \\
      + E_T K_0[E_T/T_c] (1/2) (r_1-1) t_{x1}^2\bigg\},
\end{eqnarray}
for the quark channel in the multiple transitions (MT) 
scenario.
The first term in braces comes from integrating
along the cooling curve $t = (T_0/T)^3 t_0$ from $T_0$ to
$T_c$ and the second, from the coexistence line at $T_{\chi}$,
which we approximate as being equal to $T_c$. The shortening
of the quark phase is reflected in the presence here of
$r_1=40/19$, as opposed to $r=37/3$ for the single transition
(ST) at $T_{\chi}=T_c$ case.

The corresponding rate for $\pi^{+}\pi^{-}$-annihilations 
is likewise,
\begin{eqnarray}
\label{piXsctn}
\frac{dN^{MT}_\pi}{dydM^2d E_T}= 
\pi R_A^2 A_{\pi}(M) 
   \bigg\{ 3 E_T^{-5} T_0^6 t_0^2 (G[E_T/T_0]-G[E_T/T_c])
\nonumber \\
+3 E_T^{-5} T_c^6
          t_{x2}^2 (G[E_T/T_c]-G[E_T/T_f])+
         E_T K_0[E_T/T_c] (t_{x2}^2-t_{x1}^2)/2 \bigg\}
\end{eqnarray}
\noindent
where,
\begin {eqnarray}
G[x]&=&x^3(8+x^2)K_3[x]
\end{eqnarray}
and $K_i[x]$ are the modified Bessel functions.

In this case, the first term in braces (absent altogether 
for the ST case) results from the cooling of
pions in the QGPP, from $T_0$ and $T_c$, and the second, 
from their cooling in the hadronic phase, from $T_c$ to 
$T_f$.  The latter coincides with 
the ST result if $t_{x2} \rightarrow t_2$. The last
term comes from integrating over time, 
the volume fraction, $f(t)$, of the pion phase,
on the coexistence line $T_{\chi} \simeq T_c$. 
Since pions are present both above
and below each of the coexistence lines at $T_{\chi}$ 
and $T_c$, $f(t) = 1$ for both these transitions.
In the approximation, $T_{\chi} = T_c$, the 
integral is then simply,
\begin{equation}
\int_{t_{x1}}^{t_{x2}} t dt = (t_{x2}^2 - t_{x1}^2)/2.
\end{equation}
The analogous integral in the ST case is, by contrast,
$(r-1)rt_1^2/2$.

We note that these differential rates further integrate 
over $E_T$ to the closed-form expressions:
\begin {eqnarray}
\label{integrated_qXsctn}
\frac{d N^{MT}_q}{dydM^2} =  \pi R_A^2 A_q(M)
T_0^2 t_0^2
\bigg\{6 (T_0/M)^{-4}(H[M/T_0]-H[M/T_c])
\nonumber \\
+ (M/T_c)(T_0/M)^{-4} K_1[M/T_c](r_1-1) \bigg\}
\end {eqnarray}
\noindent
where, 
\begin {equation}
H[x]=x^2(8+x^2)K_0[x]+4x(4+x^2)K_1[x]
\end {equation}
and we have translated ratios of times into ratios
of temperatures. Similarly,
\begin {eqnarray}
\label{integrated_piXsctn}
\frac{d N^{MT}_\pi}{dydM^2} = 
\pi R_A^2 A_{\pi}(M) T_0^2 t_0^2
\bigg\{ 6 (T_0/M)^{4} (H[M/T_0]-H[M/T_c])
+6 (r_1r_2)^2 (T_0/M)^{4} 
\nonumber \\    
(H[M/T_c]-H[M/T_f])
          + (M/T_c) (T_0/T_c)^4 K_1[M/T_c] [(r_1r_2)^2-1]\bigg\}
\end {eqnarray}
\noindent
Finally, the differential rates, 
$dN^{MT}_{tot}/dydM^2dE_T$ 
and $ dN^{MT}_{tot}/dydM^2$
for the {\it total} number of pairs, are  
obtained by summing the right hand sides of 
Eqs.(\ref{qXsctn}) and (\ref{piXsctn}), and 
Eqs.(\ref{integrated_qXsctn}) and (\ref{integrated_piXsctn}) 
respectively.

We can now compare our scenario, in quantitative detail,
with that of Ref.\cite{ref1}. For ease of comparison,
we use the same parameter-values: $t_0 = 1 fm/c$, and
$\pi R_A^2 = 127 fm^2$.

\begin{itemize}
\item In Fig.1, we plot 
$dN^{MT}_{tot}/dydM^2dE_T$, $dN^{MT}_q/dydM^2dE_T$ and 
$dN^{MT}_{\pi}/dydM^2dE_T$ 
{\it vs} $M$, for the combinations of $T_0$ and
$T_c$ used in Ref.\cite {ref1} but at values of $E_T$,
for which (in the first two cases) the ST pion 
peaks, coming from annihilation via the $ \rho$ 
and $\rho'$ resonances, are largely extinguished.
We note that in the MT scenario, the rate displayed 
reduces in magnitude as $E_T$ increases, but the
pion peaks survive and 
stay well above the $q\bar{q}$ contribution.
This results from pions being 
created when the plasma forms, as opposed to when it 
hadronizes.
It is consistent with the observation in \cite{ref1} that
peak extinction at large $E_T$ and the initial temperature
of the pions are distinctly correlated: the smaller this 
temperature, 
the smaller the $E_T$ at which the peaks get extinguished.

\item We note that an enhancement in 
in the production of large $E_T$ {\it dileptons}, owing to  
pions being available at  higher temperatures, will
inevitably be accompanied by an enhancement in the production of 
large $p_T$ {\it pions}, a feature highlighted by Schucraft \cite {ref14} 
long ago. We can add that pions present at the boundary
of the initial plasma, will, being colourless, have 
no problem escaping from the interaction region.
The availability of pions at higher temperatures is, 
incidentally,
supported by the observation that the HBT size of the 
initial pion source is smaller than if pions formed
only after hadronization \cite {ref15,ref16}.

\item We would intuitively expect the difference between 
the two scenarios to depend (among other factors) on the 
time spent by the system in the QGPP phase. This would in 
turn be determined by the initial temperature, $T_0$. In 
Fig.2, we accordingly plot the ratio of   
$dN_{tot}/dydM^2$ for the MT and ST scenarios {\it vs} $T_0$ 
over the range $T_{\chi} \le T_0 \le T_{comp}$,
for several values of $M$.
We find that the MT rate is $20-30\%$ higher, when $M$
is not too far from the $\rho$, $\rho'$ resonances,
but drops sharply as $M$ increases and the drooping
pion form factor begins to cut down the pion contribution.

\item We have noted that quarks acquiring constituent 
masses larger than $T_{\chi}$ and decoupling at the 
chiral phase transition,
reduces the lifetime of the quark phase {\it above} $T_c$. 
This, in turn, brings down the
$q\bar{q}\rightarrow l^+l^-$ rate for small $M$ and $E_T$.
In Fig.1, the difference is visible but only when
$T_c = 240 MeV$ and $T_0 = 250 MeV$. In the total rate,
the effect is swamped by the pion contribution 
for those values of $M$ for which the pion form
factor is appreciable. However, from Fig.3, we see
that for slightly larger values of $M$, the pion
contribution disappears and the total differential
rate $dN_{tot}/dydM^2$ reduces to just the quark
contribution, for both the ST and MT cases. 
Determining the rate in this $M$-window would
then give an indication of whether
a chiral transition has occurred close
to but distinctly above $T_c$.

\item The behaviour of these curves as functions of $T_0 $ 
can be further understood as follows:
In the ST scenario, a higher $T_0$ does not change the pion 
contribution to
dilepton production. However, it clearly increases the time 
spent by the quarks in
the cooling phase and thus enhances the quark
contribution to dileptons, pushing the crossover to the quark
dominated regime to smaller M.
For  the MT scenario, both the quark and pion contrbutions 
to dileptons
are enhanced and so the crossover to the quark-dominated 
regime is
not expected to change much (though it may still move 
slightly towards
smaller M, due to steep fall in the pion form factor).
For both scenarios, the cooling term in the quark
contribution increases in magnitude with $T_0$ while
the Maxwell co-existence term (which accounts for the
difference between the two scenarios) remains unchanged.
Thus as $T_0$ increases, the ST and MT quark
contributions approach each other.

\item Finally, in Fig.4, we plot the ratio of 
$dN_{tot}/dydM^2$ for the MT and the No Transition (NoT)
scenarios {\it vs} $T_0$. The latter corresponds to
there being no phase transition, i.e. to the fireball's
cooling in the pure hadronic phase. It is seen that
this ratio is enormous.

\item We conclude that if the dilepton rates increase
dramatically, a phase transition has probably occurred.
If the pion peaks in 
$dN_{tot}/dydM^2dE_T$ {\it vs} $M$
survive to high values of $E_T$, the 
plasma has a good chance of being a QGPP. If the 
peaks do {\it not} survive, but rather reduce to
a flat line, we are quite likely seeing a quark
phase. The actual value of this constant rate, for 
the window in $M$ discussed above, will then
decide whether the transition to the confined
hadronic phase occurred directly or through
a chirally broken phase.
\end{itemize}

In summary, we have argued that there {\it are} reasons to 
believe
that a QPG may form in heavy-ion collisions, as a result of 
several, as opposed to a single, phase transition. We have 
explored this possibility on the basis of some simple 
assumptions. We have found that this scenario pushes to 
higher temperatures the advent of the QGP, but opens up, 
in return, the exciting possibilty of a QGPP, containing 
pions 
as non-Goldstone-boson bound states, 
well above $T_{\chi}$. 
We have further investigated experimental signatures 
based on dilepton production, which
would help distinguish the single- from the 
multiple-transition scenario. Making this
distinction would shed light not merely on
how the QGP forms but on a range of assumptions
which lie at the very core of our understanding of
strong interaction dynamics.

{\bf Acknowledgement}

VS would like to thank many colleagues and collaborators,
George Ripka, Manoj Banerjee, Bojan Golli, Mike Birse, W. Broniowoski, 
J.P. Blaizot, N. D. Haridass, G. Baskaran, M.Rho and many others. VS thanks
R. Rajaraman for suggesting that an experimental signature of the 
model used here should be presented in the context of heavy ion physics. 
We apologize for missing many relevant references --- the list is too 
extensive. 

\noindent
{\bf Figure Captions}

\noindent
{\bf Figure 1} Plots of $dN^{MT}_{tot}/dydM^2dE_T$ {\it vs} 
$M$ for three $(T_0, T_c)$-combinations. For the values of
$E_T$ in Fig.1.a and Fig.1.b, the pion peaks in the single 
transition rate are essentially non-existent. Also plotted
are the quark contributions for both the MT (thick line)
and ST (thin line) scenarios.
These can be distinguished only for $T_0 = 250 MeV$ and
$T_c = 240 MeV$. The MT quark contribution is lower
because of the shortening of the quark phase.

\noindent
{\bf Figure 2} Plots of the ratio of the 
$dN_{tot}/dydM^2$
rate {\it vs} $T_0$ for the MT and ST scenarios. For the
{\it lower} $M$-values considered in these graphs,
the pion form factor is appreciable. It seen that
the MT rate is distinctly higher than the ST rate.

\noindent
{\bf Figure 3} Plots of the ratio of the 
$dN_{tot}/dydM^2$
rates for the MT and ST scenarios {\it vs} $T_0$. 
The pion contribution is essentially absent for the
{\it higher} $M$-values chosen. Note that the MT rate 
is now {\it lower} than the ST rate. The dashed lines 
represent the ratio of just the quark contributions.
Quarks contribute fewer dileptons
in the MT scenario, where the quark 
phase is shortened  by the occurrence of
the chiral transition. 

\noindent
{\bf Figure 4} The ratios of the MT rate to the rate
if no transition occurred at all.

\end{document}